\documentclass[prb,superscriptaddress,twocolumn,floatfix]{revtex4}

\usepackage{color}
\usepackage{graphicx}
\usepackage{amsmath}

\begin{document}

\title{Combined density-functional and dynamical cluster quantum Monte Carlo
calculations for three-band Hubbard models for hole-doped cuprate
superconductors}
\author{P. R. C. Kent}
\affiliation{Center for Nanophase Materials Sciences, Oak Ridge National Laboratory, Oak
Ridge, Tennessee 37831}
\author{T.Saha-Dasgupta}
\affiliation{S. N. Bose National Centre for Basic Sciences, Kolkata 700 098, India}
\author{O.~Jepsen}
\affiliation{Max-Planck-Institut f\"ur Festk\"orperforschung, D-70506 Stuttgart, Germany}
\author{O.K.Andersen}
\affiliation{Max-Planck-Institut f\"ur Festk\"orperforschung, D-70506 Stuttgart, Germany}
\author{A.~Macridin}
\affiliation{Department of Physics, University of Cincinnati, Cincinnati, Ohio 45221}
\author{T. A. Maier}
\affiliation{Center for Nanophase Materials Sciences, Oak Ridge National Laboratory, Oak
Ridge, Tennessee 37831}
\author{M.~Jarrell}
\affiliation{Department of Physics, University of Cincinnati, Cincinnati, Ohio 45221}
\author{T. C. Schulthess}
\affiliation{Center for Nanophase Materials Sciences, Oak Ridge National Laboratory, Oak
Ridge, Tennessee 37831}
\date{\today}

\begin{abstract}
Using a combined local density functional theory (DFT-LDA) and quantum Monte
Carlo (QMC) dynamic cluster approximation approach, the parameter dependence
of the superconducting transition temperature $T_{c}$ of several
single-layer hole-doped cuprate superconductors with experimentally very
different $T_{c\max }$ is investigated. The parameters of two different
three-band Hubbard models are obtained using the LDA and the downfolding $N$%
th-order muffin-tin orbital technique with $N=0$ and 1 respectively. QMC
calculations on 4-site clusters show that the $d$-wave transition
temperature $T_{c}$ depends sensitively on the parameters. While the $N$=1MTO
basis set which reproduces all three $pd\sigma $ bands leads to a $d$-wave
transition, the $N$=0 set which merely reproduces the LDA Fermi surface and
velocities does not. 
\end{abstract}

\maketitle

\section{Introduction}

Despite intense experimental and theoretical efforts, an understanding of
high-temperature superconductivity (HTSC) in hole-doped cuprate materials is
elusive. While the materials are increasingly well characterized\cite%
{HbkSCMaterBookDACardwell2003}, a firm theoretical linking of the
superconducting transition temperature $T_{c}$ to details of the underlying
atomistic and electronic structures remains a grand challenge in
condensed-matter theory.

Recent advances in quantum cluster theories have given insight into the
two-dimensional (2D) one-band Hubbard model, the most commonly adopted
many-electron model for these materials. At low temperatures and appropriate
hole concentrations, the model develops the requisite strong $%
d_{x^{2}-y^{2}} $ order and superconducting ground state\cite%
{TAMaierPRL2005,DSenechalPRL2005,SSKancharlaCONDMAT2005} due to magnetically
driven pairing\cite{TAMaierPRL2006,TAMaierPRB2006,TAMaierPRB2007}. If this
model captures sufficient physics to describe real materials, the magnetic
interactions and resultant $T_{c}$ should be moderated by details of the
actual materials and their electronic structures.

In parallel with the Hubbard investigations, the electronic structure of
HTSCs have been extensively studied using density functional theory \cite%
{WEPickettRMP1989} (DFT). Although structural properties are well
reproduced, conventional DFT-local density approximation (LDA) calculations
fail to describe the undoped insulating ground state \cite%
{JZaanenPhysicaC1988,VAnisimovPRB1991, ASvanePRL1992,WMTemmermanPRB1993}.
Nevertheless, DFT calculations agree on a universal electronic structure in
these materials: the low-energy electronic degrees of freedom are primarily
the $pd\sigma $ anti-bonding O$\,p$ and Cu$\,d_{x^{2}-y^{2}}$ orbitals in
the CuO$_{2}$ layer, and these bands have been parameterized \cite%
{OKAndersenJPCS1995,AKMcMahanPRB1990,MSHybertsenPRB1989,FMilaPRB1988}. For
all optimally and over-doped materials, the Fermi surfaces (FSs) measured by
angle-resolved photoemission (ARPES)\cite{ADamascelliRMP2003} agree
surprisingly well with detailed LDA predictions. Even the existence and
distinct in-plane dispersion of the splitting between the two FS sheets in
bi-layered cuprates \cite{OKAndersenJPCS1995} and the $k_{z}$-dispersion in
body-centered tetragonal single-layered materials\cite{EPavariniPRL2001},
have recently been experimentally confirmed\cite%
{ADamascelliRMP2003,NEHusseyNat2003}. Finally, the LDA conduction-band
parameter $r\sim -t^{\prime }/t,$\cite{OKAndersenJPCS1995} which gives the
material-dependence of the FS shape and has the same origin as $t_{\perp
}\left( \mathbf{k}_{||}\right) $ has been found to correlate positively with 
$T_{c\max }$ \cite{EPavariniPRL2001}, but causal links have not been
established.

In this paper we combine LDA-DFT calculations of the cuprates with quantum
cluster calculations of the transition temperatures $T_{c}$ of the 2D
Hubbard model. We study the three- rather than the one-band model because
the most localized Cu $d_{x^{2}-y^{2}}$-like orbital describing the LDA
conduction band is so extended,\cite{OKAndersenPRB2000} that using merely
the on-site Coulomb repulsion in a one-band model is not justified. However,
including also the O$_{x}\,p_{x},$ and O$_{y}\,p_{y}$ orbitals in the basis
set, localizes the Cu $d_{x^{2}-y^{2}}$ orbital to the extent that the
corresponding three-band Hubbard Hamiltonian appears to be a valid model.
DFT calculations are used to consistently obtain the parameters for the five
single-layer materials HgBa$_{2}$CuO$_{4}$, Tl$_{2}$Ba$_{2}$CuO$_{6}$,
TlBaLaCuO$_{5}$, La$_{2}$CuO$_{4}$, and Ca$_{2}$CuO$_{2}$Cl$_{2},$ for which 
$T_{c\,\max }=90\,$K, $85\,$K, 52$\,$K, 40$\,$K, and 26$\,$K$,$respectively.
To determine $T_{c}$ we use the dynamic cluster approximation (DCA) with a
finite-temperature quantum Monte Carlo (QMC) cluster solver \cite%
{MJarrellPRB2001,MHHettlerPRB1998,MHHettlerPRB2000} and the previously
calculated DFT parameters\cite{TSahaDasguptaUnpub}. In principle, these
calculations use no experimental input and are therefore a stringent test of
both the density functional and quantum cluster methods, as well as the form
of the underlying model. As the first study of this type we aim to address
the following questions: (1) What is the magnitude of $T_{c}$ variation in
the Hubbard model following the LDA+DCA scheme, and is this variation
realistic? (2) Are there parameters found by LDA-DFT beyond those typically
considered in Hubbard-like schemes that are particularly important for
determining $T_{c}$ in these materials\cite{QYinPRL2008}.

\section{Density functional calculations}

\label{sec:dft}

\begin{figure*}[tbp]
\includegraphics*[width=4.0in,keepaspectratio]{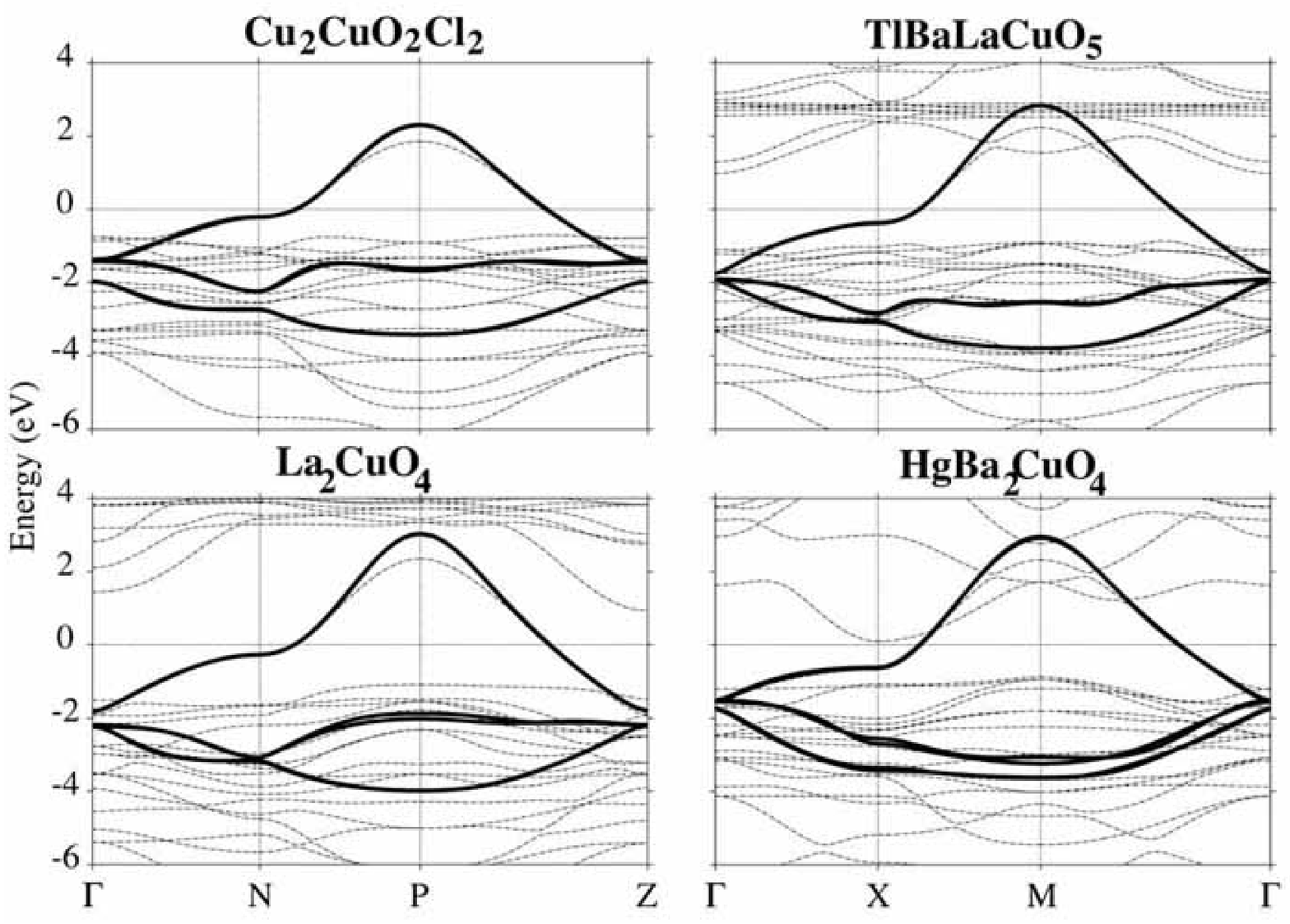}
\caption{Comparison of the downfolded N=0 MTO 3-band (bold) and complete LDA
bandstructures (dashed) of four undoped single-layer HTSC materials. The
zero of energy is the Fermi level. }
\label{fig:bstruc}
\end{figure*}

We approximate the LDA potential for the stoichiometric (undoped) cuprates
by a superposition of spherically-symmetric, overlapping potential wells and
then construct the basis set of three orbitals per cell by downfolding
within multiple-scattering theory \emph{at} the LDA Fermi energy, $\epsilon
_{F}$. Such an orbital is constructed to have the following properties: (1)
It solves Schr\"{o}dinger's differential equation at $\epsilon _{F}$
throughout the solid, i.e. in all partial-wave channels, except for a kink
at the muffin-tin (MT) spheres in the Cu$\,d_{x^{2}-y^{2}},$ O$_{x}\,p_{x},$
and O$_{y}\,p_{y}$ channels. (2) It has \emph{no} Cu$\,d_{x^{2}-y^{2}},$ O$%
_{x}\,p_{x},$ or O$_{y}\,p_{y}$ character inside any MT sphere other than
the one in which the orbital is centered and has its own character. Hence,
the orbital is chosen to vanish (with a kink) in the channels of the other
orbitals, and this makes it maximally localized. Pictures are presented in
Ref.$\,$\cite{TSahaDasguptaUnpub}. This basis set of kinked partial waves 
\cite{OKAndersenPRB2000} provides Bloch solutions of Schr\"{o}dinger's
equation with errors proportional to $\varepsilon \left( \mathbf{k}\right)
-\epsilon _{F},$ and energy bands with errors proportional to $\left[
\varepsilon \left( \mathbf{k}\right) -\epsilon _{F}\right] ^{2}$ due to the
variational principle. The LDA FS and velocities are thus given correctly,
as can clearly be seen from Fig.$\,$\ref{fig:bstruc}. For the solutions of
Schr\"{o}dinger's equation at $\epsilon _{F}$, the kinks cancel out.

\begin{figure*}[tbp]
\includegraphics*[width=4.0in,keepaspectratio]{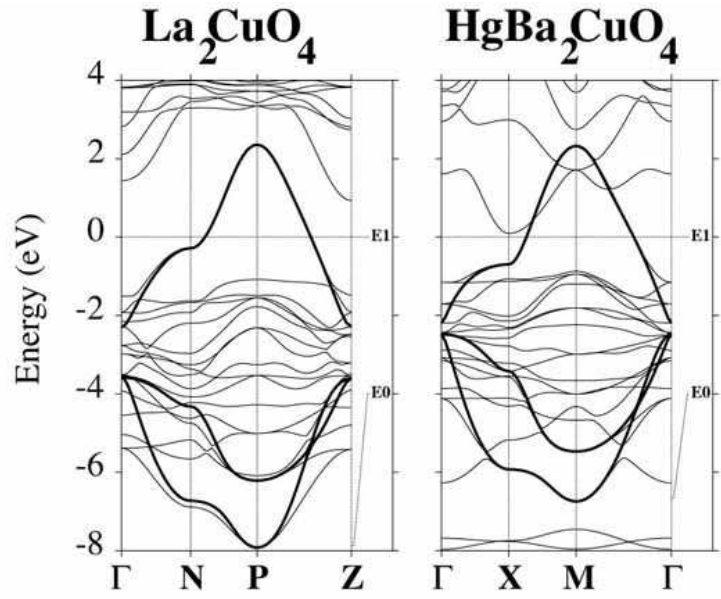}
\caption{Comparison of the downfolded N=1 3-band (bold) and complete LDA
bandstructures (dashed) of the single-layer HTSC materials HgBa$_{2}$CuO$%
_{4} $ and La$_{2}$CuO$_{4}$. The zero of energy is the Fermi level. }
\label{fig:bstrucemery}
\end{figure*}

Instead of using kinked partial waves at $\epsilon _{F}$ as basis functions,
we could have constructed $N$th-order muffin-tin orbitals ($N$MTOs) \cite%
{OKAndersenPRB2000} which for a \emph{mesh} of energies, $\epsilon
_{0},..,\epsilon _{N},$ yield wavefunctions with errors proportional to $%
\left[ \varepsilon \left( \mathbf{k}\right) -\epsilon _{0}\right] ..\left[
\varepsilon \left( \mathbf{k}\right) -\epsilon _{N}\right] $. In that way,
the three bands can be made to reproduce the LDA bands over a wider energy
range, specifically the range set by $U_{dd}\approx 10\,$eV. In addition to
the above-mentioned $N\mathrm{=}0$ set$,$ we shall also consider an $N%
\mathrm{=}1$ set with the second energy, $\epsilon _{1},$ chosen near the
bottom of the $pd\sigma $ bonding band, i.e. $7-8$ eV below $\epsilon
_{0}\equiv \epsilon _{F}.$ As seen in Fig. \ref{fig:bstrucemery}, this basis
set accounts for the LDA $pd\sigma $ bonding, non-bonding, and antibonding
bands, at the same time as it reproduces the LDA FS and velocities. However,
its orbitals are slightly less localized.

Symmetrical orthormalization of the three $N$MTOs finally yields the orbital
representation in which $\hat{H}$ is expressed. The on-site elements of $H_{%
\mathrm{LDA}}$ are the orbital energies, $\epsilon _{\alpha }^{\mathrm{LDA}%
}, $ and the off-site elements are the integrals for hopping, $t_{\alpha
ij}^{\beta lm},$ between orbital $\alpha $ on site $ij$ and orbital $\beta $
on site $lm$ in the same CuO$_{2}$ layer. Inter-layer hopping is neglected
in this study. The site indices for the $d$ orbital are integers, since it
is on a cubic lattice, while those for the $x$ $\left( y\right) $ orbital
are $\frac{1}{2}0$ $\left( 0\frac{1}{2}\right) $ plus integers. The
orbitals, centered on the Cu and O sites, are orthonormal and real by
construction. On-site energies and hopping integrals are, therefore, real
and symmetric. The Fourier components of $H(\mathbf{k})_{\mathrm{LDA}}$ are,
for example $H_{dd}(\mathbf{k})=\epsilon _{d}+2t_{d00}^{d10}(\cos k_{x}+\cos
k_{y})+4t_{d00}^{d11}\cos k_{x}\,\cos k_{y}+\ldots $. 
Table \ref{tab:notation} gives the short notation used for the hopping integrals.

\begin{table*}
\begin{tabular}{ccccccccccc}
$t_{d000}^{d100}$ & $t_{d000}^{d110}$ & $t_{d000}^{x\frac{1}{2}00}$ & $t_{d000}^{x\frac{1}{2}10}$ & $t_{d000}^{x\frac{3}{2}00}$ & $t_{d000}^{x\frac{3}{2}10}$ & $t_{x\frac{-1}{2}00}^{y0\frac{1}{2}0}$ & $t_{x\frac{-1}{2}00}^{x\frac{1}{2}00}$ & $t_{x\frac{1}{2}00}^{x\frac{1}{2}10}$ & $t_{x\frac{-1}{2}00}^{x\frac{1}{2}10}$ & $t_{x\frac{-1}{2}00}^{y1\frac{1}{2}0}=t_{x\frac{-1}{2}00}^{y0\frac{3}{2}0}$ \\ 
$t_{dd}$ & $t_{dd}^{\prime }$ & $t_{pd}$ & $t_{pd}^{\prime }$ & $t_{pd}^{\prime \prime }$ & $t_{pd}^{\prime \prime \prime }$ & $t_{pp}$ & $t_{pp}^{\prime }$ & $t_{pp}^{\prime \prime }$ & $t_{pp}^{\prime \prime \prime }$ & $t_{pp}^{\prime \prime \prime \prime }$
\end{tabular}
\caption{Relationship between site index and short-form notation used for hopping integrals}
\label{tab:notation}
\end{table*}

\begin{figure*}[tbp]
\includegraphics*[width=2.5in,keepaspectratio,angle=-90]{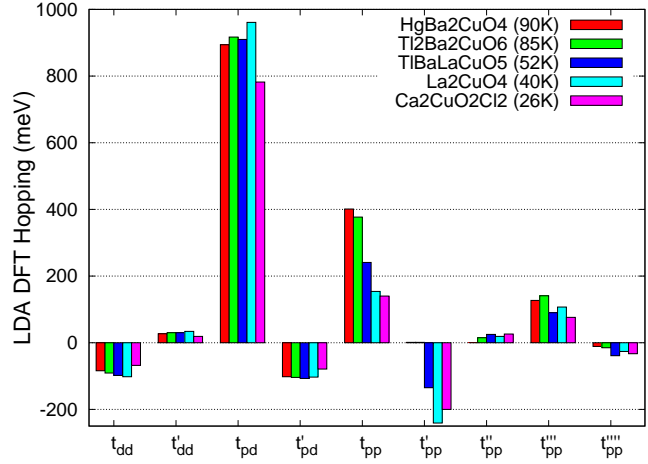}
\caption{(Color online) Calculated in-layer hoppings for five single-layer
cuprates for N=0. The materials are ordered by $T_{c\max }.$}
\label{fig:ldahopp}
\end{figure*}

The values of the hopping integrals for the $N=0$ basis set are shown in Fig.$\,$\ref{fig:ldahopp}
 They have reasonably short range and are
dominated by the usual $t_{pd}$ and $t_{pp}.$
The values $\sim 0.9\,$eV of $t_{pd}$ are considerably smaller than the
conventional value 1.5$\,$eV \cite%
{OKAndersenJPCS1995,AKMcMahanPRB1990,MSHybertsenPRB1989,FMilaPRB1988}
describing the width $\sim 4\sqrt{2}t_{pd}\sim 8.5\,$eV of the $pd\sigma $
anti-bonding \emph{and} bonding bands. Whereas $t_{pd}$ and most other
hopping integrals are seen to be fairly independent of the material (only
the one with apical-Cl is a bit smaller), $t_{pp}$ is \emph{not;} it
increases with the observed $T_{c\,\max }.$ This is the conduction-band
trend found previously \cite{EPavariniPRL2001} and explained as $p_{x}$ to $%
p_{y}$ hopping via a high-energy $\left( \epsilon _{s}\right) ,$
Cu-centered, \emph{axial} hybrid consisting of Cu$\,4s,$ Cu 3$d_{3z^{2}-1},$
apical oxygen $2p_{z}$ and axial cation orbitals, all stacked perpendicular
to the layer. If the energy of this axial orbital is increased (e.g. by
moving apical oxygen closer to Cu), $t_{pp}\sim \frac{t_{sp}^{2}}{\epsilon
_{s}-\epsilon _{F}}$ decreases \cite{OKAndersenJPCS1995}. Within that axial
model, $t_{pp}^{\prime }=t_{pp},$ but Fig.$\,$\ref{fig:ldahopp} shows that
this is not true: $t_{pp}^{\prime }$ vanishes for the two high-$T_{c}$
cuprates, and for the three low-$T_{c}$ cuprates the sign of $t_{pp}^{\prime
}$ is opposite to that of $t_{pp}$. The main reason is that hopping via the
in-layer (and therefore material-independent) Cu $4p_{x}$ orbital
contributes to $t_{pp}^{\prime },$ but not to $t_{pp},$ and opposes the
hopping via the axial orbital\cite{TSahaDasguptaUnpub}. Also
material-independent hopping via O$_{y}\,p_{x}$ orbitals influences $%
t_{pp}^{\prime }$, and causes a sizeable $t_{pp}^{\prime \prime \prime }.$
So $t_{pp}$ and $t_{pp}^{\prime }$ exhibit the material's trend. Finally, $%
t_{dd}$ proceeds mainly via the diffuse O$_{x}$ $3d_{3x^{2}-1}$ orbital,
lying 50 eV above $\epsilon _{F},$ and $t_{pd}^{\prime }$ proceeds mainly
via polarization of the cation. In summary, (1) diffuse high-energy $\left(
\epsilon _{\gamma }\right) $ orbitals make sizeable contributions $\propto
t_{\alpha \gamma }^{2}/\left( \epsilon _{F}-\epsilon _{\gamma }\right) $ to
the LDA hopping integrals and (2) the energy $\epsilon _{s}$ of the axial
orbital, here downfolded into the tails of the oxygen orbitals, is the
essential material-dependent parameter.

The Cu on-site Coulomb energy is $U_{dd}\approx 9.5\,$eV in all five
cuprates as found by constrained LDA \cite{OGunnarssonPRB1989} calculations
with the LMTO-ASA method. The radius of the Cu sphere is adjusted to $1.32\,$%
\AA , such that the $d$-character in the upper half of the LDA conduction
band, the $d$ hole-count $h_{d}^{\mathrm{LDA}},$ is the same as that
obtained from the three-band $H_{\mathrm{LDA}}.$ This ensures that the Cu$%
\,d_{x^{2}-y^{2}}$ partial wave truncated outside the atomic sphere is
similar to the Cu$\,d_{x^{2}-y^{2}}$ partner of the three-orbital $N$MTO
set. For the four cuprates with apical oxygen, we find: $h_{d}^{\mathrm{LDA}%
}=0.46$ and for the diagonal elements of $H_{\mathrm{LDA}}$: $\epsilon _{d}^{%
\mathrm{LDA}}\left( h_{d}\right) -\epsilon _{p}\approx 0.45\,$eV. Since
correlation effects are already taken into account at the mean-field level
in the LDA, we must include a double-counting correction proportional to the
deviation, $h_{d}^{\mathrm{LDA}}-\frac{1}{2},$ of the $d$ hole-count from
that $\left( \frac{1}{2}\right) $ of the $d^{9}\rightarrow d^{10}$
transition-state. The corrected orbital-energy difference is then: $\Delta
\equiv e_{p}-e_{d}\,=\,\epsilon _{d}+U_{dd}-\epsilon _{p}\,=\,\epsilon _{d}^{%
\mathrm{LDA}}(h_{d}^{\mathrm{LDA}})-\epsilon _{p}+(h_{d}^{\mathrm{LDA}}-%
\frac{1}{2})U_{dd}\,\approx \,0.0\,$eV, where $e$ refers to the hole and $%
\epsilon $ to the electron representation. The commonly assumed value is,
however, $\Delta \sim 3\,$eV.\cite%
{AKMcMahanPRB1990,MSHybertsenPRB1989,FMilaPRB1988} Previous constrained LDA
calculations for La$_{2}$CuO$_{4}$ \cite{MSHybertsenPRB1989,AKMcMahanPRB1990}
gave 3$\,$eV because they used the LMTO-ASA total $d$ electron-count of 9.24
to deduce: $h_{d}^{\mathrm{LDA}}=0.76,$ and then found the double-counting
correction to be $\sim $2.5\thinspace eV. 
However, integrating to the top of the conduction band, we find not 10, but
9.70 $d$ electrons, which is consistent with $h_{d}^{\mathrm{LDA}}=0.46$. 
\cite{PRCKentFootNote2007b}

Unfortunately, current many-body treatments fail to reproduce the insulating
behavior at half-filling, unless $\Delta $ exceeds 3$\,$eV, and this is
commonly felt to be unacceptable. We therefore empirically set $\Delta
=3.25\,$eV in $H_{ \mathrm{LDA}},$ i.e. we increased $\epsilon _{d}^{\mathrm{%
LDA}}-\epsilon _{p} $ by 3.25 eV, but kept our hopping integrals unchanged.%
\cite{PRCKentFootNote2007a} This does not completely ruin the agreement
between the experimental and LDA FS shapes, but it weakens the trend: For Tl$%
_{2}$Ba$_{2}$CuO$_{6}$ the effective $-t^{\prime }/t$ is reduced from 0.33 
\cite{EPavariniPRL2001} to 0.22, with the experimental\cite{NEHusseyNat2003}
value being 0.28, while for La$_{2}$CuO$_{4}$ the reduction is merely from
0.17 to 0.16. In all five cases, the $\Delta $-shift causes a 20\% reduction
of the effective conduction bandwidth $8t$.%

With the $N\mathrm{=}1$ basis set, which describes the three LDA $pd\sigma $
bands over the energy range $U_{dd},$ rather than merely the antibonding
band near the LDA Fermi level (see Fig.s \ref{fig:bstruc} and \ref%
{fig:bstrucemery}), the values of the hopping integrals for HgBa$_{2}$CuO$%
_{4}$ (90$\,$K) and La$_{2}$CuO$_{4}$ (40$\,$K) are as shown in Fig.\ref%
{fig:ldahoppemery}. Now $t_{pd}$ is increased to values much closer to the
conventional ones \cite%
{OKAndersenJPCS1995,AKMcMahanPRB1990,MSHybertsenPRB1989,FMilaPRB1988} and $%
t_{pp}$ is increased to 0.90 eV. By having to span a wider energy-range, the 
$N\mathrm{=}1$ orbitals are somewhat less localized, and consequently have
somewhat longer-ranged hoppings, than the $N\mathrm{=}0$ orbitals. This also
masks the material's trend in individual hopping integrals, although it is
of course present in the shape of the antibonding band near $\varepsilon _{F}
$. Now $\epsilon _{d}^{\mathrm{LDA}}-\epsilon _{p}=0.67$ and 0.95 eV for HgBa%
$_{2}$CuO$_{4}$ (90$\,$K) and La$_{2}$CuO$_{4},$ respectively, but for the
reason mentioned above, we shall set $\Delta $ to 3.25 eV.

\section{Dynamic Cluster Approximation Calculations}

\label{sec:dca}

To solve the 3-band Hubbard Hamiltonian $\hat{H}$ we use the DCA\cite%
{MHHettlerPRB2000,MHHettlerPRB1998} (for a review, see Ref. \cite%
{TAMaierRMP2005}). In this method we map the lattice model onto a periodic
cluster of size $L_{c}\times L_{c},$ embedded into a self-consistently
determined mean-field background. Correlations up to a range $\mathrm{\sim }%
L_{c}$ are treated explicitly while longer-ranged correlations are treated
at a mean field level. We solve the cluster problem using QMC\cite%
{MJarrellPRB2001}, which does not introduce further significant
approximations. Calculations on large clusters at low temperatures become
prohibitively (exponentially) expensive due to the QMC Fermion sign problem. 
$T_{c}$ is determined via the diverging $d$-wave pair-field susceptibility
obtained over a series of calculations at progressively lower temperatures.
We check for earlier divergences in other angular momentum channels.

Due to the large computational cost of a parametric study using QMC,
we have performed calculations on 4-site clusters ($L_{c}\mathrm{=}2$)
at 15\% hole doping, which is near optimal in real materials. The
4-site cluster is the smallest for which a $d$-wave order parameter is
allowed topologically, and corresponds to a mean-field result
\cite{TAMaierPRL2005}. Single-band calculations on 4-site clusters
have shown that the phase diagram of these clusters shows general
agreement with HTSC\cite{TAMaierRMP2005}. Converged calculations on
clusters of up to 26 sites - for a single set of parameters - find
that $T_{c}$ of 4-site clusters is \emph{over}-estimated by a factor
$%
\sim $2,\cite{TAMaierPRL2005} i.e. the small clusters exhibit larger
pairing correlations. Hence, the presence of $d$-wave order in 4-site
clusters at low temperature does not confirm the existence of such
order in larger clusters, while the absence of $d$-wave order strongly
indicates an absence of this order in the thermodynamic limit. In all
of our DCA calculations we consistently used the calculated hoppings $t$ and not,
for example, the original LDA dispersion.

To establish the existence of a variation in $T_{c}$ in the three-band model
we performed an initial parametric study using only the nearest-neighbor
hopping integrals $t_{pd}$ and $t_{pp}$. Leaving one of these fixed at the
value calculated with the $N\mathrm{=}0$ basis set for HgBa$_{2}$CuO$_{4},$
our highest $T_{c\max }$ material, we varied the other. As shown in Fig.\ref%
{fig:2ptc}, a $d$-wave transition was obtained over the entire range of
parameters studied. Increasing either parameter increased $T_{c}$, and $%
dT_{c}/dt_{pd}\sim 6.3\times dT_{c}/dt_{pp}$. The increase of $T_{c}$ with
increased $t_{pd}$ may be understood in terms of changes to the fundamental
energy scale. Within the range of studied materials, however, the calculated
variation of $t_{pd}$ is rather insignificant due to very similar Cu-O bond
lengths while $t_{pp}$ varies systematically, with larger values
corresponding to materials with larger $T_{c\max },$ a trend reproduced by
our QMC calculations.

\begin{figure*}[tbp]
\includegraphics*[width=4.0in,keepaspectratio]{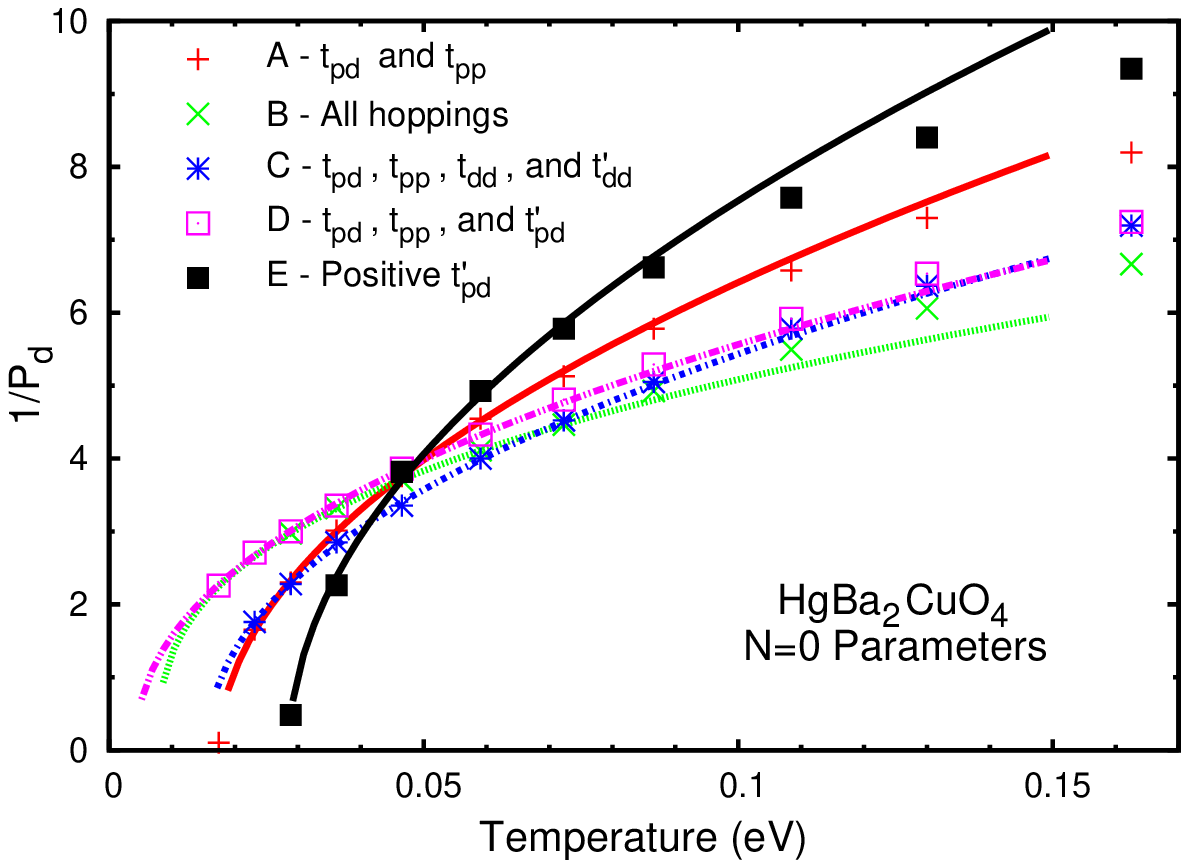}
\caption{(Color online) Inverse d-wave pair-field susceptibility ($%
P_{d}^{-1} $)as a function of temperature for different hopping-parameter
sets calculated from HgBa$_{2}$CuO$_{4}$ with $N\mathrm{=}0$ (see text). The
lines are power law fits to the lowest 5 points in each series. }
\label{fig:fullptc}
\end{figure*}

In Fig.\ref{fig:fullptc} we show the calculated inverse $d$ pair-field
susceptibility as a function of temperature for HgBa$_{2}$CuO$_{4}$ with the 
$N\mathrm{=}0$ basis set when all hoppings are included, as well as subsets
of hoppings. Compared to the inverse susceptibility of the $t_{pd}$-$t_{pp}$
only calculation (A), which yields a moderate $T_{c}\sim 17\,$meV, when
including all the hoppings (B) the inverse susceptibility is reduced at high
temperatures, but reduces less quickly at lower temperatures. Therefore, any
transition for the true LDA $N\mathrm{=}0$ hoppings must occur at much lower
temperatures than for the $t_{pd}$-$t_{pp}$ only case. The $d$ hole
occupancy at low temperature is $h_{d}\sim $0.76 compared to $\sim $0.78 for
the $t_{pd}$-$t_{pp}$ case. For the complete set (B) there appears to be no $%
d$-wave transition at moderate temperatures. Due to the increasing
computational cost for lower temperatures we cannot completely exclude the
possibility of a very-low-temperature transition, but it is certain that any 
$T_{c}$ is significantly reduced from the simpler $t_{pd}$-$t_{pp}$ only
case.When calculations are performed for all five materials (not shown), we
also find no apparent $d$-wave transitions. These results demonstrate that,
surprisingly, $T_{c}$ of the three-band Hubbard model is a strong function
of the hopping parameters beyond the nearest-neighbors.

To investigate the cause of the $T_{c}$ reduction we systematically surveyed
the effect of varying each hopping parameter to extract $dT_{c}/dt_{\alpha
ij}^{\beta lm}$. In (C) we see that adding $t_{dd}$ and $t_{dd}^{\prime }$
to $t_{pd}$ and $t_{pp}$ is not what suppresses $T_{c}$, but adding $%
t_{pd}^{\prime }$ does, as seen in (D). Although all hopping parameters
modify $T_{c}$, variation of $t_{pd}^{\prime }$ changes $T_{c}$ most
dramatically. This hopping proceeds mainly via polarization of the cation
and is $-0.10\,$eV for all five materials. Changing the sign of $%
t_{pd}^{\prime },$ but keeping all other hoppings realistic, even produces a
significant enhancement of $T_{c}$ as shown in (E). This artificial sign
change profoundly changes the $U\mathrm{=}0$ conduction band: the effective $%
t$ decreases by a factor 2 from the full LDA value, and $-t^{\prime }/t$
decreases from 0.34 to 0, i.e. this change is opposite to the emperical trend%
\cite{EPavariniPRL2001}.

\begin{figure*}[tbp]
\includegraphics*[width=2.5in,keepaspectratio,angle=-90]{%
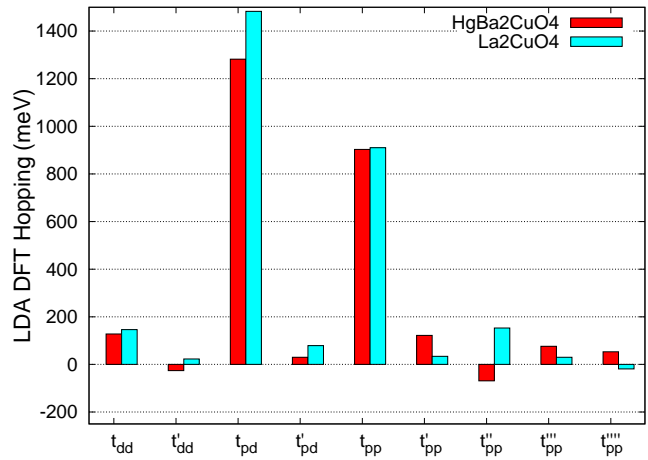}
\caption{(Color online) Comparison of calculated in-layer hoppings for two
single-layer cuprates for N=0 and N=1 (see text)}
\label{fig:ldahoppemery}
\end{figure*}

We now repeat the calculations using the $N\mathrm{=}1$ basis set which
reproduces all three LDA $pd\sigma $ bands over a range of 10 eV $\sim U_{dd}
$ (see Fig.s \ref{fig:bstrucemery} and \ref{fig:ldahoppemery}). The inverse $%
d$ pair-field susceptibilities for HgBa$_{2}$CuO$_{4}$ are shown in Fig. \ref%
{fig:emeryptc}. When all the hoppings are included (B), the susceptibility
is increased at all temperatures compared to the $N\mathrm{=}0$ case, and in
contrast to the previous results, there does appear to be a $d$-wave
transition at very low temperature. This result clearly demonstrates a
strong sensitivity of the many-body results on details of the treatment of
the LDA data and Hubbard Hamiltonian.

\begin{figure*}[tbp]
\includegraphics*[width=4.0in,keepaspectratio]{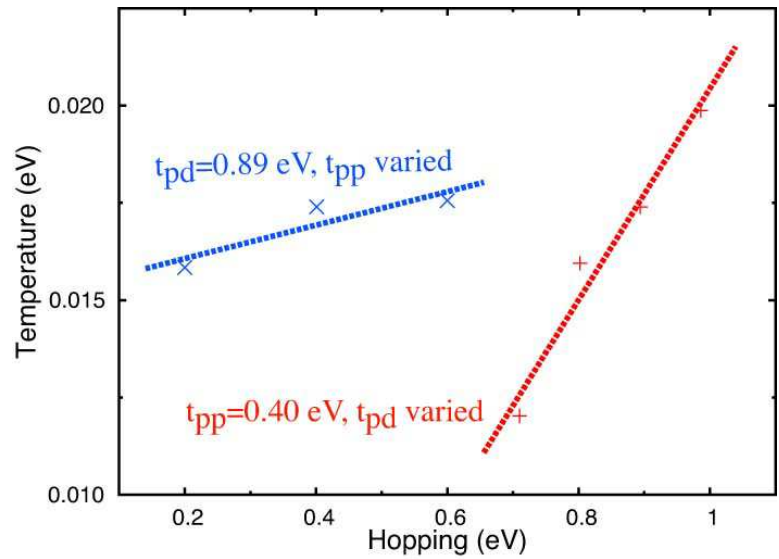}
\caption{(Color online) Calculated transition temperatures $T_{c}$ for the
3-band $t_{pd}$-$t_{pp}$ model (see text). The temperature variation is
shown for variations in $t_{pp}$ with $t_{pd}$, held fixed (blue crosses),
and vise versa (red plusses). The guide lines are a linear fit.}
\label{fig:2ptc}
\end{figure*}

\begin{figure*}[tbp]
\includegraphics*[width=4.0in,keepaspectratio]{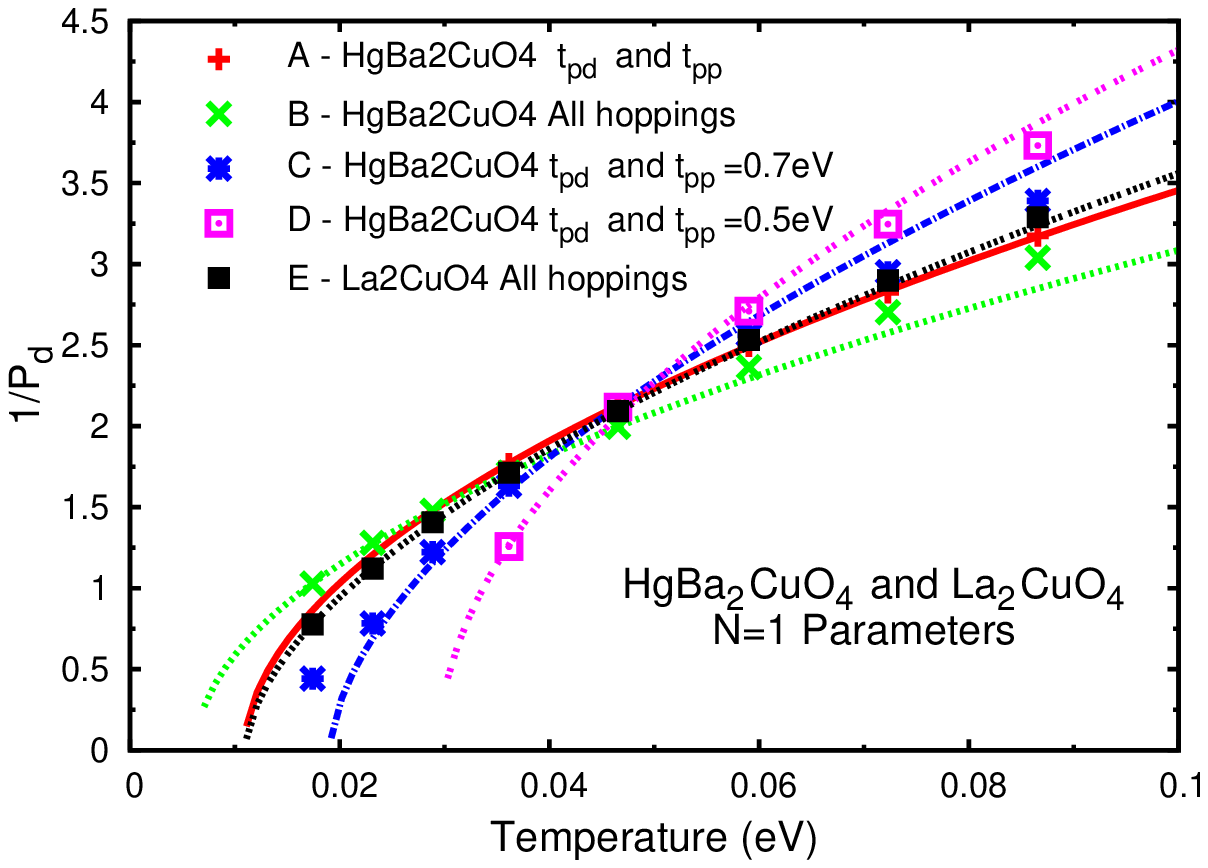}
\caption{(Color online) Inverse d-wave pair-field susceptibility ($%
P_{d}^{-1} $)as a function of temperature for different hopping-parameter
sets calculated from HgBa$_{2}$CuO$_{4}$ and for La$_{2}$CuO$_{4}$ with $N\mathrm{=}1$ (see text). The
lines are power law fits to the lowest 5 points in each series. }
\label{fig:emeryptc}
\end{figure*}

To further investigate the differences between the $N\mathrm{=}0$ and
$N%
\mathrm{=}1$ parameter sets, we performed $N\mathrm{=}1$ calculations
including only $t_{pp}$ and $t_{pd}$. From Fig.\ref{fig:emeryptc} we
see that neglecting all hoppings except these two between nearest
neighbors has little effect on the susceptibility (A). This result is
in marked contrast with the one for the $N\mathrm{=}0$ set, where the
removal of long-ranged terms, particularly $t_{pd}^{\prime }$,
significantly increased $T_{c}$.  This is presumably connected with
the fact that $t_{pd}^{\prime }$ changes sign when going from
$N\mathrm{=}0$ to $N\mathrm{=}1$. We also show the effect of
\emph{decreasing} $t_{pp}$ with $t_{pd}=1.28\,$eV (C,D): in contrast
to results for $t_{pd}=0.89$ eV (Fig.\ref{fig:2ptc}) we find $T_{c}$
to \emph{%
  increase. }This does not contradict the emperical trend that
$T_{c\,\max }$ increases with $t^{\prime }/t$ because this does not
translate into an increase with $t_{pp}$ for the $N\mathrm{=}1$
set. Comparisons of the inverse susceptibility for calculations
performed with the full parameters sets for HgBa$_{2}$CuO$_{4}$ (B)
and La$_{2}$CuO$_{4}$ (E) indicates that La$_{2}$CuO$_{4}$ will have
the higher transition temperature, the reverse order compared to experiment.

As an additional independent test on the choice of downfolding method
we repeated the QMC calculations using HgBa$_{2}$CuO$_{4}$ hopping
parameters determined by a Wannier-function projection
method\cite{WKuPRL2002} which, like the $N\mathrm{=}1$ set, reproduced
all three $pd\sigma $ bands.  Nevertheless, in this case we found
\emph{no} $d$-wave transition in the computationally accessible
temperature range (a low temperature transition cannot be ruled out).

The above results clearly demonstrate that the phase diagram of the
three-band Hubbard model is quite sensitive to the choice of hopping
integrals, even around the commonly accepted energy range of $t_{pd}\sim 1$%
eV. Although a $d$-wave transition is found for the $N\mathrm{=}1$MTO basis
set which reproduces all three $pd\sigma $ LDA bands as well as the LDA FS
and velocities, no transition is found for slightly different choices of
downfolding approach, e.g. for Wannier-function projection of the three
bands or for the $N\mathrm{=}0$MTO basis which only reproduces the LDA FS
and velocities. We have only investigated a single point on the phase
diagram due to the computational expense and numerical difficulty of the
current QMC and DCA techniques. Within the three-band Hamiltonian,
refinement of the ill-determined $\Delta =e_{p}-e_{d}$ is clearly required,
as well as investigation of the effect of different hoppings on the spectral
properties. It is also highly desirable to investigate larger clusters as
well as more complex Hamiltonians: for example, in LDA-DFT the $%
d_{3z^{2}-r^{2}}$ band lies close to the Fermi energy in some of the HTSC
materials, suggesting that additional Cu degrees of freedom may be required.
Unfortunately these investigations are currently precluded due to the
computational cost and worsening Fermion sign-problem; we hope they will be
examined in future.

\section{Conclusions}

\label{sec:conclusions}

In summary, we have obtained the parameters of three-band Hubbard models for
a series of single-layer cuprate superconductors with varying $T_{c\,\max }$
from downfolding either to the LDA conduction band or to all three $pd\sigma 
$ bands. The transition temperature calculated using DCA-QMC on 4-site
clusters and increasing the small LDA value of $e_{p}-e_{d}$ to 3.25$\,$eV
is a moderate to very strong function of the hopping parameters. Even small
hopping integrals beyond the nearest-neighbours can have a marked effect on
the transition temperatures. These parameters are sensitive to the choice of
downfolding technique and effective degree of Wannier localization. The
present calculations yields superconductivity for the NMTO basis set which
reproduces all three $pd\sigma $ LDA bands, but not for the one which
reproduces merely the LDA Fermi surface and velocities. We hope that our
findings will motivate future investigations and methodological development
of more robust approaches for constructing and/or for solving realistic
models of the cuprate superconductors.

We thank O. Gunnarsson, I. Dasgupta, J.P.\ Hague and D.J.\ Scalapino for
useful discussions, and W. Ku for providing alternative hopping parameters.
TSD and OKA acknowledge the MPG-India partner group program. A portion of
this research was conducted at the Center for Nanophase Materials Sciences
at Oak Ridge National Laboratory, used computational resources of the Center
for Computational Sciences, and was sponsored by the offices of Basic Energy
Sciences and Advanced Scientific Computing Research, U.S. Department of
Energy. AM and MJ were supported by CMSN DOE DE-FG02-04ER46129 and NSF
DMR-0312680.

\begin{thebibliography}{30}
\bibitem[{PRC({\natexlab{a}})}]{PRCKentFootNote2007b}
\bibinfo{note}{The 0.3 $d$ electron missing is $d_{3z^{2}-1},$ which extends in
  the $z$-direction, beyond the sphere. In our three-band model, this orbital
  is downfolded into the tails of the O$\,p$-orbitals
  \cite{TSahaDasguptaUnpub}.}
\bibitem[{PRC({\natexlab{b}})}]{PRCKentFootNote2007a}
\bibinfo{note}{Using a two-hopping parameter model we find $T_c$ variatons of
  1\% for $e_p-e_d=3.5$ and $4.79$ eV.}
\end{thebibliography}
\end{document}